\documentclass[aps,pre,reprint,groupedaddress]{revtex4-2}%
\usepackage{graphicx,pstricks}
\usepackage{amsmath,amssymb,amsfonts}
\usepackage{amsthm}%
\usepackage{xcolor}
\usepackage{paralist}

\newcommand{\nn}{\nonumber}\newcommand{\vphi}{\varphi}
\newcommand{\veps}{\varepsilon}\newcommand{\ea}{eqnarray}
\newcommand{\eas}{eqnarray*}
\newcommand{\tl}{\tilde}

\def\bexa{\begin{example}}\def\eexa{\end{example}}
\def\brem{\begin{remark}}\def\erem{\end{remark}}
\def\bthm{\begin{theorem}}\def\ethm{\end{theorem}}
\def\blem{\begin{lemma}}\def\elem{\end{lemma}}
\def\bcor{\begin{corollary}}\def\ecor{\end{corollary}}
\def\bdefi{\begin{definition}}\def\edefi{\end{definition}}

\def\bmip{\begin{minipage}{\textwidth}}\def\emip{\end{minipage}}
\def\huga#1{\begin{gather} #1 \end{gather}}
\def\hugast#1{\begin{gather*} #1 \end{gather*}}
\def\hual#1{\begin{align} #1 \end{align}}
\def\hualst#1{\begin{align*} #1 \end{align*}}

\newcommand{\btab}[2]{\begin{tabular}{#1}#2\end{tabular}}

\def\sm{\small}\def\ig{\includegraphics}\def\tew{\textwidth}

\def\ds{\displaystyle}
\def\lam{\lambda}
\newcommand{\bi}{\begin{itemize}}\newcommand{\ei}{\end{itemize}}
\newcommand{\ben}{\begin{enumerate}}\newcommand{\een}{\end{enumerate}}
\newcommand{\bce}{\begin{center}}\newcommand{\ece}{\end{center}}
\newcommand{\bci}{\begin{compactitem}}\newcommand{\eci}{\end{compactitem}}
\newcommand{\bcen}{\begin{compactenum}}\newcommand{\ecen}{\end{compactenum}}
\newcommand{\bcena}{\begin{compactenum}[(a)]}
\newcommand{\reff}[1]{(\ref{#1})}

\newcommand{\nonu}{\nonumber}

\newcommand{\hs}[1]{{\hspace{#1}}}\newcommand{\vs}[1]{{\vspace{#1}}}

\newcommand{\barr}{\begin{array}}\newcommand{\earr}{\end{array}}
\newcommand{\bpm}{\begin{pmatrix}}\newcommand{\epm}{\end{pmatrix}}
\newcommand{\bsm}{\left(\begin{smallmatrix}}
\newcommand{\esm}{\end{smallmatrix}\right)}

\def\er{{\rm e}}

\def\sig{\sigma}

\def\bd{\begin{displaymath}} \def\ed{\end{displaymath}}

\begin{document}
\title{Electrolyte flows under magnetic fields: Manning-like counterion condensation in one 
dimension}
\author{Yoav Tsori}
\email{tsori@bgu.ac.il}
\affiliation{Department of Chemical Engineering, Ben-Gurion University of the 
Negev, Beer-Sheva, Israel}  
\author{Hannes Uecker}\email{hannes.uecker@uol.de}
\affiliation{Institute for Mathematics, Carl von Ossietzky Universit\"at Oldenburg, Oldenburg, 
Germany} 
\date{\today}

\begin{abstract}
We present a theoretical framework for unidirectional electromagnetohydrodynamic flow 
of dilute electrolytes under perpendicular magnetic fields. Starting from the Navier--Stokes 
equation coupled with the Poisson--Nernst--Planck formulation, we show that the problem 
admits a sequential decoupling: the Stokes equation is solved first to obtain the 
velocity profile, which defines a hydrodynamic potential entering the Nernst--Planck description of 
ions. This Lorentz-force-induced potential competes with electrostatic attraction and significantly 
alters ionic distributions. We analyze this mechanism in two canonical geometries. In 
planar Couette shear, it produces a Manning--Oosawa-like condensation transition in 
one dimension, a phenomenon absent in classical electrostatics. We derive an eigenvalue 
equation predicting a sharp threshold between counterion enrichment and depletion at the 
charged wall. In cylindrical Taylor--Couette flow, the same effect shifts the classical 
Manning criterion by a magnetic parameter, enabling tunable control of condensation. These 
findings extend Manning--Oosawa phenomenology to driven, non-equilibrium systems and 
provide a basis for magnetic manipulation of screening in electrolytes, with 
implications for microfluidics, electrochemical systems, and nonlinear boundary-value 
theory.
\end{abstract}

\maketitle


\section{Introduction}\label{isec}
Magnetohydrodynamics (MHD) couples the Navier-Stokes equations of
fluid motion with Maxwell’s equations of electromagnetism, leading to
a rich set of nonlinear phenomena. It plays a pivotal role in plasma
confinement in fusion devices \cite{poedts_book}, solar and
astrophysical processes \cite{arregui_book,priest_book}, and geodynamo
action in Earth’s core \cite{king_rpp_2013}. Its applied importance
spans energy conversion (MHD generators) \cite{rosa_book},
metallurgical processing \cite{matinde_jsaimm_2024}, electromagnetic
pumps and emerging microfluidic technologies
\cite{bau_magnetochemistry_2022,nguyen_mn_2012a,nguyen_mn_2012b}. Despite
its broad utility, MHD remains mathematically challenging due to the
strong coupling between flow and electromagnetic fields, motivating
both analytical and computational research \cite{davidson_book}.

MHD of electrolytes is traditionally formulated in terms of momentum transport with 
electromagnetic body forces and Ohmic conduction, typically assuming bulk electroneutrality and 
treating the ionic distribution as uniform 
\cite{muller_book,fahidy_jae_1983,nguyen_mn_2012a,nguyen_mn_2012b}.
Standard macroscopic MHD strictly assumes bulk electroneutrality and neglects electrostatic 
forces. In large systems, ionic densities are essentially equal to each other and to their common 
bulk value, matching this assumption. However, near charged interfaces, on the scale of the Debye 
length, electroneutrality breaks down. In this regime, the volumetric Lorentz force includes strong 
electrical contributions, and the flow is more accurately described as an 
electromagnetohydrodynamic (EMHD) flow 
\cite{sarkar_microfluid_2017,chen_colloids_2018,rana_colloids_2022}.

We present a formulation of one-dimensional EMHD for dilute
electrolytes in unidirectional flow that explicitly captures this boundary layer 
\cite{bazant_prfluids_2016}: the Lorentz force acting on mobile ions induces a {\it hydrodynamic
  potential} that couples the Poisson-Nernst-Planck system to the
Stokes flow in a way that allows a sequential decoupling of the
hydrodynamics from the electrostatics. Crucially, while the ionic density differences are confined to 
the micro-scale, the Lorentz force drives a macroscopic variation in the electrostatic potential. 

Consequently, once the unidirectional velocity profile is solved from the Stokes equation, it 
generates a hydrodynamic potential whose competition with electrostatic attraction
determines the steady counterion distribution via a modified Poisson-Boltzmann (PB) equation. 

Here we apply the new formulation to simple, analytically
tractable geometries. We focus on counterion-only systems and analyze
(i) planar Couette shear and (ii) concentric, infinitely long coaxial
cylinders (Taylor-Couette flow without instabilities). Despite the
simplicity of these flows, we uncover a Manning–Oosawa (MO)--like
condensation transition in both cases, driven not by electrostatics
alone but by the bulk magnetic forcing.

Manning-Oosawa condensation describes the partial neutralization
of a highly charged cylindrical macroion by its counterions
\cite{manning_jcp_1969,oosawa_book}. Mathematically, it is a
peculiarity of the two-dimensional (2D) logarithmic Coulomb potential:
in infinite systems, there exists a critical line-charge parameter
above which a finite fraction of counterions \emph{condense} near the
macroion \cite{kornyshev_prx_2024}, while below it they \emph{run
  away} to infinity
\cite{netz_joanny_mm_1998,burak_orland_pre_2006,netz_orland_pre_2000,trizac_prl_2006}. In 
contrast, 
in one-dimensional (1D) planar geometry, counterions are
always bound (no condensation threshold), whereas in three-dimensional
(3D) spherical geometry counterion entropy wins over electrostatic
attraction and they escape (no bound state)
\cite{trizac_prl_2002,trizac_prl_2011,andelman_ariel_epl_2003}. Thus, the appearance of a 
condensation threshold is a hallmark of 2D
electrostatics and does not occur in 1D within the classical, purely
electrostatic PB description
\cite{andelman_chapter_2006,levin_mm_1998}.

Within our EMHD-PB framework, the magnetic field and flow jointly
generate a hydrodynamic potential that supplements the electrostatic
potential in the Boltzmann weight. This additional, spatially varying
term, competes with the electrostatic attraction and can qualitatively change
binding. We show that even in planar, one-dimensional shear (a
geometry that classically lacks a condensation threshold), there
emerges a sharp criterion separating a regime with finite counterion
density at the wall from a regime in which counterions escape to
infinity, i.e., an MO-like transition \emph{in 1D}. In cylindrical
geometry, the same mechanism leads to a shift of the classical Manning
threshold by a dimensionless magnetic parameter.

These sharp criteria apply in the limit of infinite flows, i.e., 
when for instance the distance $d$ of the walls confining the flows goes 
to infinity. For finite systems, there is a continuous transition 
between counterions bound to the wall and counterions driven away from 
the wall, occurring at some finite strength $B=B_c\sim 1/d^2$. We 
shall analyze in detail the behaviour of the ions distribution $n^+(x)$ 
for $B$ near $B_c$.

Section~II presents the EMHD formulation and the emergence of a
hydrodynamic potential in unidirectional flows. Section~III analyzes
planar Couette shear flow and establishes the magnetic-field–induced
MO-like transition in one dimension, including the eigenvalue
condition and near-critical behavior. Section~IV treats the
coaxial-cylinder geometry, deriving the shifted Manning criterion. 
Section~V concludes with implications and prospects.

\section{EMHD of dilute electrolytes in unidirectional flow}
\label{mhdsec}
We consider the flow of incompressible dilute electrolytes past one or two
surfaces under the influence of a magnetic field. The Navier-Stokes
equation is \cite{probstein_book}
\begin{\ea}
  \rho\left[\frac{\partial {\bf v}}{\partial t}+({\bf v}\cdot{\bf
      \nabla}){\bf
      v}\right]&=&-\nabla p+\eta\nabla^2{\bf v}+\rho_e{\bf E}\nn\\
  &&+{\bf j}\times{\bf B}-k_BT\nabla n^+;
\label{eq_navier_stokes}
\end{\ea}
${\bf v}$ is the liquid's velocity, $\rho$ its mass density,
$n^+$ the counterion number density, $p$ the pressure, $\eta$
the liquid viscosity, ${\bf E}$ and ${\bf B}$ the electric and
magnetic fields, respectively, and ${\bf j}$ is the electric current
density.

The terms with ${\bf E}$ and ${\bf B}$ originate from the
Lorentz force experienced by an ion of charge $q$,
${\bf f_{\rm L}}=q\left({\bf E}+{\bf v}\times {\bf B}\right)$, which
transfers to the liquid. We assume a salt-free electrolyte where the only mobile charges are the 
positive counterions released from the moving wall; we denote the local charge
density as $\rho_e=qn^+$. The term $k_BT\nabla n^+$ corresponds to the osmotic pressure
of an ideal counterion gas, where $k_B$ is the Boltzmann constant and $T$
the absolute temperature.

\begin{figure}[ht!]
  \centering 
\includegraphics[width=0.3\textwidth,bb=125 50 425 520,clip]{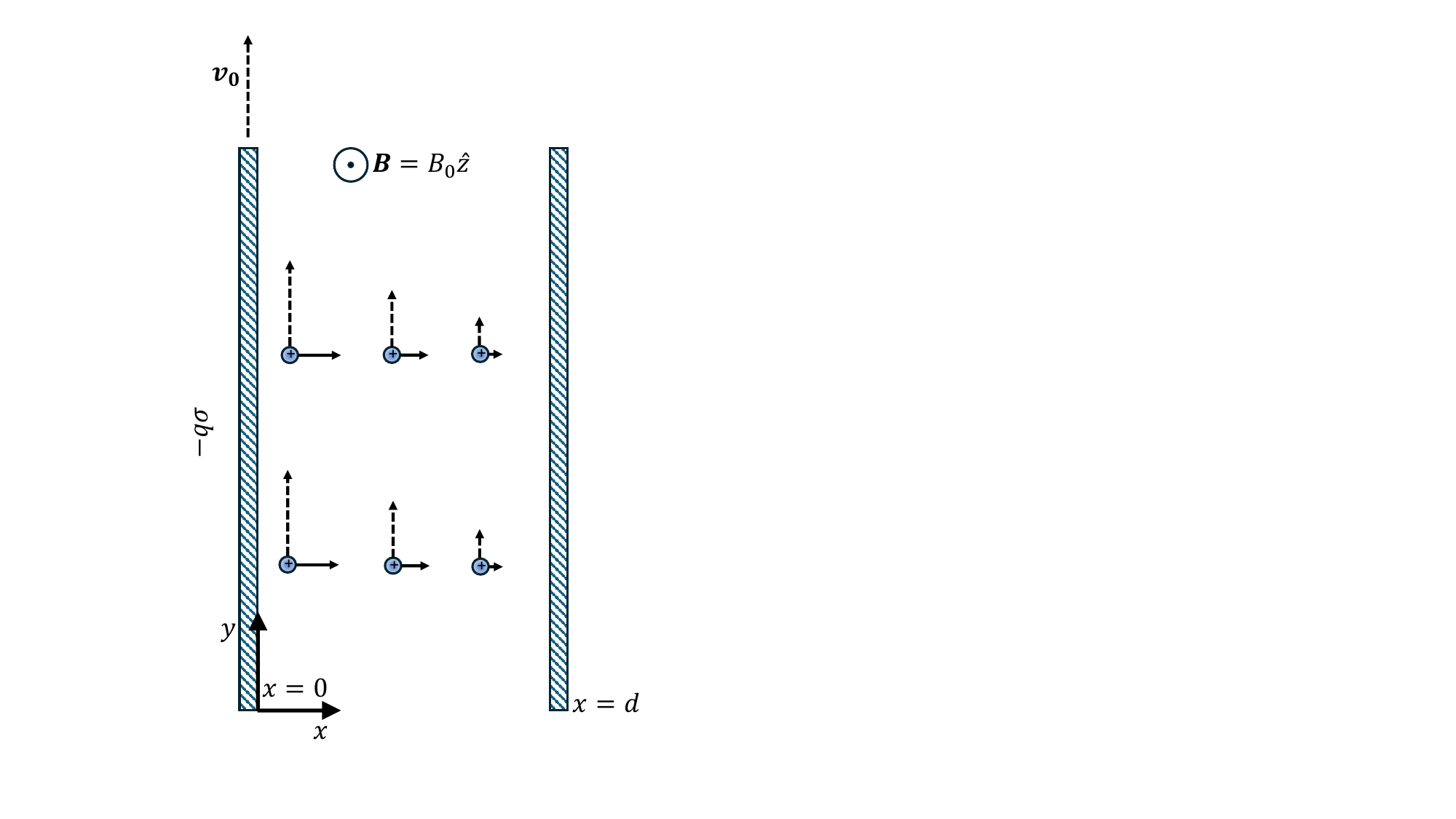}
  \caption{System illustration. A liquid is confined between a surface
    at $x=0$ ($-q\sigma$ charge density) moving at a speed $v_0$ in
    the $y$ direction and a stationary surface at $x=d$. The liquid
    moves in unidirectional flow along $y$, dragging the positive
    counterions. An external magnetic field is applied in the $z$
    direction. Dashed arrows illustrate the local velocity; solid
    rightward arrows illustrate the Lorentz force
    ${\bf f}_L=q\left({\bf E}+{\bf v}\times{\bf B}\right)$ acting on
    counterions. }
  \label{fig1_illus}
\end{figure}

The total electrical current density is fundamentally the sum of the individual ionic fluxes, ${\bf j} = 
\sum q_i {\bf h}_i$, when there are several ionic species. In standard bulk MHD, this reduces to the 
classical Ohm's law, ${\bf j} = 
\sigma_e({\bf E} + {\bf v}\times {\bf B})$. However, to capture the non-electroneutral boundary 
layer, 
we must retain the explicit Poisson-Nernst-Planck formulation. The ionic flux of counterions has 
three 
contributions: Lorentz force, diffusion, and
advection with the liquid, 
\begin{\ea} {\bf h}&=&n^+ \mu q\left({\bf E}+{\bf v}\times {\bf
      B}\right)-D\nabla n^++n^+{\bf v}, 
\end{\ea}
where $\mu$ is the ionic mobility in the liquid, $D=\mu k_BT$ is the
diffusion constant, and ${\bf h}$ satisfies the Nernst-Planck equation
\begin{\ea}\label{eq_np}
  \partial_t n^++\nabla\cdot {\bf h}=0.
\end{\ea}
Our task is to solve \reff{eq_np} simultaneously with 
(\ref{eq_navier_stokes}).

We focus on unidirectional flows such as the one illustrated in
Fig.~\ref{fig1_illus}: A wall at $x=0$ is charged with $-q\sigma$
charge per unit area and moves vertically at a velocity
$v_0$. $\sigma$ is the surface charge number density. The wall at
$x=d$ is stationary and uncharged.  The fluid velocity is then
${\bf v}=v(x)\hat{y}$. In a constant perpendicular magnetic field,
${\bf B}=B_0\hat{z}$, and in the absence of electrochemistry at the
confining walls, ${\bf j}$ is parallel to $\hat{y}$, and both
${\bf E}$ and ${\bf j}\times{\bf B}$ are parallel to $\hat{x}$. In
addition, because we assume a unidirectional velocity field of the form ${\bf v}=v(x)\hat{y}$, the 
nonlinear advection term $({\bf v}\cdot{\bf \nabla}){\bf v}$ vanishes identically, regardless of the 
Reynolds number. The Navier-Stokes equations for
the $y$ component becomes the Stokes equation:
\begin{\ea}\label{eq_stokes}
\rho\partial_t
v(x)=-\partial_yp+\partial_x(\eta\partial_xv(x)).\label{ns_y}
\end{\ea}
Hence, the hydrodynamic problem in the $y$ direction is the {\it same
  as if there were no magnetic and electric fields}. Consequently,
once the flow profile $v(x)$ is found from (\ref{ns_y}), it can be
substituted in (\ref{eq_np}) to find the ionic density $n^+(x)$.

We continue in steady state, where all time derivatives vanish. When
the walls are impenetrable to ions ($h_x=0$ at the walls), the ion
flux in the $x$-direction vanishes, namely
\begin{\ea}\label{zero_flux}
  n^+ \mu q\left(E + v B_0\right)-D\partial_x
  n^+=0,
\end{\ea}
where we used ${\bf E}=E\hat{x}$ and $({\bf v}\times {\bf B})_x = v B_0$. In this problem, therefore, 
the 
ions' distribution is Boltzmann-like 
\begin{\ea}\label{eq_ions_boltzmann}
  n^+(x)&=&n_m\exp\left(\frac{-q\psi+ U_{\rm hyd}(x)}{k_BT}\right), 
\end{\ea}
where $n_m\equiv n^+(x=0)$ is the ion density at the moving wall (to
be determined) and
\begin{\ea}\label{eq_Uhyd}
  U_{\rm hyd}(x)&\equiv &q\int_0^x v(x')B_0dx'
\end{\ea}
is a hydrodynamic potential. For simple liquids and on the continuum
level \cite{kornyshev_jpcb_2008}, the electrostatic potential $\psi$
satisfies the Poisson equation $\nabla^2\psi=-qn^+/\veps$, where
$\veps$ is the dielectric constant of the liquid. Using the ion
distribution in (\ref{eq_ions_boltzmann}), we obtain a modified
Poisson-Boltzmann equation 
\begin{\ea}\label{eq_pb}
  \nabla^2\psi=-\frac{qn_m}{\veps}\exp\left(\frac{-q\psi+ U_{\rm
        hyd}(x)}{k_BT}\right).
\end{\ea}
In the next section we turn to solve equations (\ref{eq_stokes}),
(\ref{eq_Uhyd}), and (\ref{eq_pb}) in planar Couette flow.

\section{Manning-Oosawa-like condensation in Couette shear flow}\label{sec_shear}

Due to the sequential decoupling between flow profile and ion density,
in the Couette shear flow illustrated in Fig.\ref{fig1_illus}, we
readily find that
\hugast{
{\bf v}(x)=v_0\left(1-\frac{x}{d}\right)\hat{y} 
\text{ and }
  U_{\rm hyd}(x)=qv_0B_0\left(x-\frac12\frac{x^2}{d}\right).
}
The Poisson-Boltzmann equation and its boundary conditions are then
\hualst{
  \psi''&=-\frac{qn_m}{\veps}\exp\left(-\frac{q\psi}{k_BT}+\frac{qv_0B_0d}{k_BT}
    \left(\frac{x}{d}-\frac12\frac{x^2}{d^2}\right)\right),\\
  \psi'(0)&=\frac{q\sigma}{\veps}~,~~\psi'(d)=0.
}
It is instructive to use the dimensionless magnetic field $\tl{B}$ and
Bjerrum length defined as 
\hugast{
  \tl{B}=qv_0B_0d/k_BT, \quad 
  l_B= q^2/(\veps k_BT),
}
and measure distances in units $l_B$, such that $\tl{x}=x/l_B$ and
$\tl{d}=d/l_B$. Lastly, we define an auxilliary potential as
$\phi=q\psi/k_BT-\tl{B}\left(\tl{x}/\tl{d}-\tl{x}^2/(2\tl{d}^2)\right)$,
leading to
\hual{
\label{eq_shear}
  \phi''&=-\lambda e^{-\phi}+\frac{\tl{B}}{\tl{d}^2}, 
\quad\lam=l_B^3n_m, \\
  \phi'(0)&=l_B^2\sigma-\frac{\tl{B}}{\tl{d}}~,\quad \phi '(\tl{d})=0, 
\quad \phi(0)=0, \nonu
}
where, without loss of generality, we set $\phi(0)=0$ because 
any constant shift in $\phi$
simply rescales $\lambda$, so fixing $\phi(0)=0$ ensures a unique nonlinear 
``eigenvalue'' $\lam$, which encodes the unknown $n_m$. The problem at hand is a nonlinear 
second-order ordinary differential
equation with a constant ``forcing'' term, and one ``additional'' BC 
(three instead of two) 
compensated by the eigenvalue $\lam$. The subtraction of the term proportional to $\tl{B}$ in the 
definition
of $\phi$ removes the explicit $x$-dependence in the exponential,
simplifying the Poisson–Boltzmann equation to a form where the
magnetic forcing appears as a constant term. Note that 
$\ds\lam=\frac 1 2 l_B^2/\lambda_D^2$ if the Debye length $\lambda_D$ is defined
via $\lambda_D^2=\veps k_BT/(2n_mq^2)$, and that \reff{eq_shear} yields the 
conservation integral
\huga{\label{c1}
  \int_0^{\tl{d}}\lambda e^{-\phi}d\tl{x}=l_B^2\sigma.
}

\begin{figure}[ht]
  \centering
\btab{l}{
\hs{3mm}\ig[width=0.42\tew]{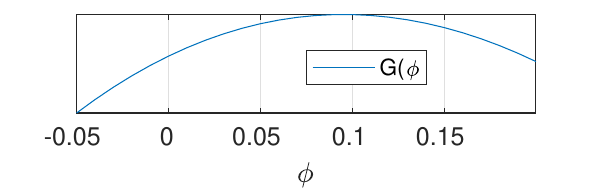}\\
\ig[width=0.44\tew]{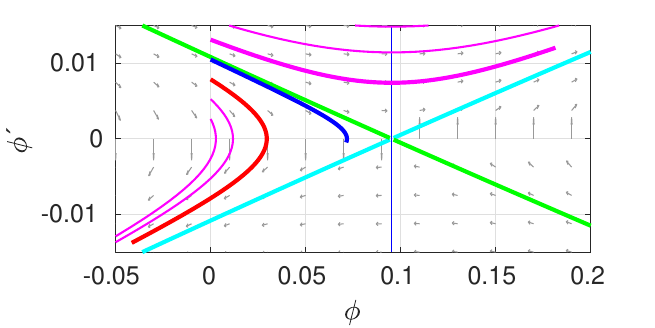}\\
\hs{3mm}\ig[width=0.4\tew]{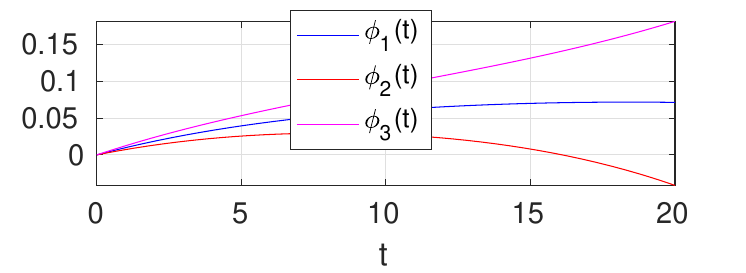}
}

\vs{-3mm}
\caption{{\sm Potential energy $G$, phase portrait, and selected solution curves for 
\reff{eq_shear}, $\tl{B}=5, \tl{d}=20$, $\lam=1.1\lam_*$. The 
vertical blue line is at the fixed point 
$\phi_*=\ln(\lam/\lam_*)$, and the green/cyan 
lines illustrate the associated stable/unstable manifolds. We integrate \reff{eq_shear} to $x=\tl{d}$ 
with $\phi(0)=0$ and varying 
$\phi'(0)$. The BC $\phi'(\tl{d})=0$ means 
that we must land on the $\phi'=0$ axis, which approximately holds for the 
blue curve. For illustration, we also show two ``nearby'' curves; the red 
curve reaches $\phi'=0$ too early, and the (thick) magenta curve too late. Even for this relatively 
small 
$\tl{d}$, the ``correct'' blue curve reaches  
$\phi'=0$ close to the fixed point $\phi_*$, hence with $\phi''\approx 0$, 
see \reff{a1}, \reff{eq_f_of_lam}.}
  \label{hf1}}
\end{figure}

\begin{figure}[ht]
  \centering
\btab{l}{
\hs{3mm}\ig[width=0.42\tew]{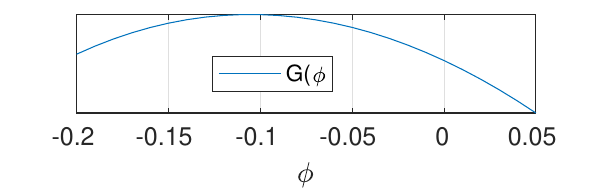}\\
\ig[width=0.44\tew]{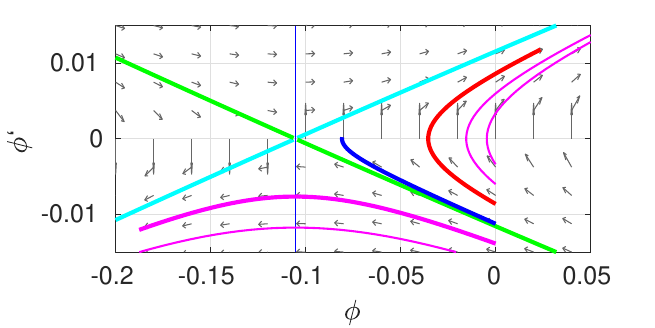}\\
\hs{3mm}\ig[width=0.4\tew]{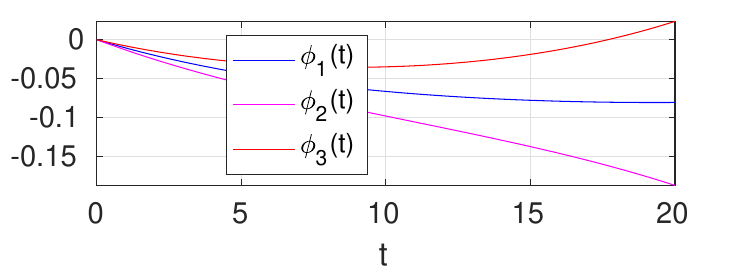}
}

\vs{-3mm}
\caption{{\sm Like Fig.\ref{hf1}, but $\lam=0.9\lam_*$. Hence fixed 
point at $\phi_*=\ln(\lam/\lam_*)<0$, and initial conditions are 
chosen with $\phi'(0)<0$.}
  \label{hf1b}}
\end{figure}

We start with a phase plane analysis 
of \reff{eq_shear}, based on the first order formulation 
\huga{\label{1st} 
\text{$y_1'=y_2,\quad y_2'=-\lam e^{-y_1}+\tl{B}/\tl{d}^2$}, 
} 
with ``time'' $\tl{x}$ and conserved energy 
\huga{\label{e1}
E=\frac 1 2 \phi'^2+G(\phi), \quad G(\phi)=-\lam\er^{-\phi}-\frac{\tl{B}}{\tl{d}^2}\phi.
}
From \reff{c1} we have $\lam>0$, and the potential energy $G(\phi)$ has a unique 
maximum at 
\huga{
\phi_*=\ln(\lam/\lam^*)~~\left\{\barr{l}>0,\quad \lam>\lam^*,\\
<0,\quad \lam<\lam^*,\earr\right.\quad \lam^*=\tl{B}/\tl{d}^2.
}
Thus, for $\phi'(0)=l_B^2\sigma-\tl{B}/\tl{d}>0$ we must have 
$\lam>\lam^*$, and for $\phi'(0)<0$ 
we obtain $0<\lam<\lam^*$, where the value of $\lam$ follows  
equivalently from $\phi'(\tl{d})=0$ or from \reff{c1}. For $\phi'(0)=0$ we have 
the unique solution $\phi \equiv 0$ with $\lam=\lam^*$. See Fig.\ref{hf1} ($\lam>\lam_*$) and  
Fig.\ref{hf1b} ($\lam<\lam_*$) 
for illustration, 
where contrary to the role of $\lam$ as an eigenvalue depending on 
$\tl{B},\tl{d}$, and $\phi'(0)$, 
we a priori fix $\lam$ near $\lam_*$ and then 
vary $\phi'(0)$.

\begin{figure}[ht!]
  \centering
\ig[width=0.4\tew]{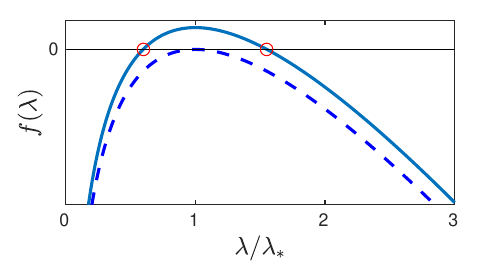}
  \caption{{\sm Illustration of the approximate eigenvalue equation
    $f(\lambda)=0$, with $f(\lambda)$ from
    (\ref{eq_f_of_lam}). Dashed curve: 
    $l_B^2\sigma= \tilde{B}/\tilde{d}$; Solid curve: 
 $l_B^2\sigma\ne \tilde{B}/\tilde{d}$, with two roots 
    $\lambda_1<\lam_*<\lambda_2$ (red circles).\label{fig_f_of_lambda}}}
\end{figure}

Since $G(\phi)$ has a unique maximum, 
$\phi(\tl{x})$ must approach $\phi(\tl{d})$ monotonically as $\tl{x}\to\tl{d}$. This implies that for 
the 
first order formulation \reff{1st} and large $\tl{d}$, 
the solution $\phi(\tl{x})$ must be close to the stable manifold of the saddle point 
$(\phi,\phi')=(\phi_*,0)$, 
which reaches $ (\phi_*,0)$ in infinite ``time'' $\tl{x}$. In particular, 
$\phi(\tl{d})\approx\phi_*=\ln(\lam/\lam_*)$, and this 
allows us to approximate $\lam$ via conservation of energy $E$. Namely, 
\hual{
0=&E(\tl{x}{=}0)-E(\tl{x}{=}\tl{d})\nonu\\
\approx&\frac12\left(l_B^2\sigma-\frac{\tl{B}}{
      \tl{d}}\right)^2-\lam +\lam_*+\lam_*\ln(\lam/\lam_*), \label{a1} 
}
which yields an approximate (in the limit $\tl{d}\to\infty$ with fixed 
$\tl{B}/\tl{d}$) equation for $\lambda$, namely 
\huga{\label{eq_f_of_lam}
  f(\lambda):=\lambda^*\left[1{+}\ln\left(\frac{\lambda}{\lambda^*}
    \right){-}\frac{\lambda}{\lambda^*}\right]+\frac12\left(l_B^2\sigma{-}\frac{\tl{B}}{
      \tl{d}}\right)^2{=}0.
}
Figure \ref{fig_f_of_lambda} shows the shape of $f(\lam)$; for 
$l_B^2\sigma=\tl{B}/\tl{d}$ we have a double root at $\lam=\lam_*$, 
while for $l_B^2\sigma\ne \tl{B}/\tl{d}$ we have two roots $\lam_1<\lam_*<\lam_2$.
Again, from the phase portraits in Fig.~\ref{hf1} we see that for 
$\phi'(0)=l_B^2\sigma-\tl{B}/\tl{d}>0$, the correct one  
is $\lam_2>\lam_*$ because then $\phi_*=\ln(\lam/\lam_*)>0$, while 
for $l_B^2\sigma-\tl{B}/\tl{d}<0$ the correct one 
is $\lam_1<\lam_*$. When $l_B^2\sigma=\tl{B}/\tl{d}$, the solution is
$\phi\equiv 0$, and as $l_B^2\sigma-\tl{B}/\tl{d}$ decreases from positive to negative
values, the potential at $\tl{x}=\tl{d}$ switches from
$\phi(\tl{d})=\ln(\lam_2/\lam_*)>0$ to 
$\phi(\tl{d})=\ln(\lambda_1/\lam_*)<0$.
The critical value of the normalized density
\begin{\ea}
  (l_B^2\sigma)_c=\frac{\tl{B}}{\tl{d}}=\frac{qv_0B_0l_B}{k_BT}
\end{\ea}
remains constant in the $\tl{d}\to\infty$ limit. However, in this limit the
critical value of $\lambda$ tends to zero:
\begin{\ea}
  \lambda^*=\frac{\tl{B}}{\tl{d}^2}=\frac{qv_0B_0l_B^2}{k_BTd}\to 0.
\end{\ea}
For the counterion density 
$n^+(\tl{x}){=}l_B^{-3}\lambda e^{-\phi(\tl{x})}$, the above yields the following: In a finite
system, there is always a finite value of $\lambda$, hence $n^+(0){=}l_B^3\lam$ at
the left wall is finite. There is a
critical value of the surface charge, $(l_B^2\sigma)_c=\tl{B}/\tl{d}$ where 
$\phi$ switches from increasing ($l_B^2\sigma>\tl{B}/\tl{d}$) to decreasing 
 ($l_B^2\sigma>\tl{B}/\tl{d}$). Physically,
it marks the point where the Lorentz-driven hydrodynamic potential
exactly balances the electrostatic attraction. When
$l_B^2\sigma<\tl{B}/\tl{d}$, the magnetic part of the Lorentz force is
strong and counterions are depleted from the wall. In the limit $\tl{d}\to\infty$, with fixed 
$\tl{B}/\tl{d}$,
 $\lam_*=\tl{B}/\tl{d}^2\to 0$, 
and for $l_B^2\sigma<(l_B^2\sigma)_c$ this implies 
that essentially $\lambda=0$, i.e., the counterion density at
the wall vanishes, and the solution corresponds to counterions
escaping to infinity.
In the classical MO condensation, a 
counterion run-off threshhold occurs only in two-dimensions and not in
one (counterions are always bounded) or three dimensions (counterions
alway run off to infinity). Here, the transition occurs in one
dimension due to the bulk (volume) force of magnetic origin. Moreover, irrespective whether 
$l_B^2\sigma-\tl{B}/\tl{d}$ is positive or
negative, from $\phi(\tl{d})\approx \phi_*=\ln(\lam/\lam_*)$ the counterion density at
$\tl{x}=\tl{d}$ is proportional to $\tl{B}/\tl{d}^2$ and tends to zero
as $\tl{d}\to\infty$.
To estimate the critical value of $\sigma$ we use $q=e$,
$v_0=0.1$m/s, $B_0=0.1$T, $l_B\approx 10$ nm and room
temperature. One then finds a low critical charge density
$(l_B^2\sigma)_c=\tl{B}/\tl{d}\approx 4\times 10^{-9}$ indicating that
usually a finite fraction of counterions will be bound to the surface.
\begin{figure}[ht]
  \centering
\btab{l}{
\btab{l}{{\sm (a)}\\
\ig[width=0.4\tew]{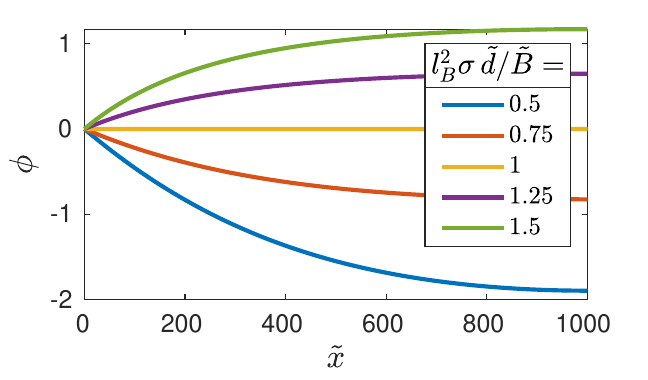}\\
{\sm (b)}\\
\ig[width=0.4\tew]{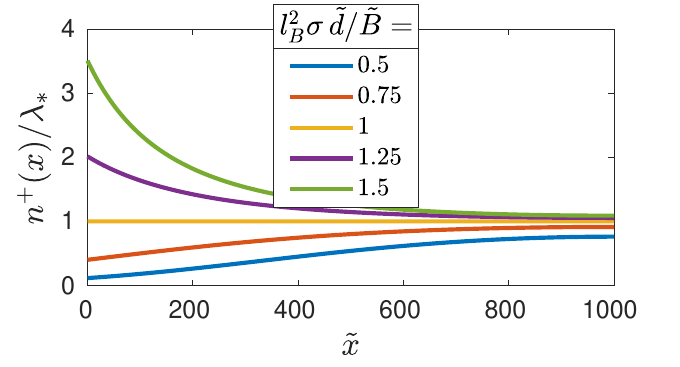}}\\
{\sm (c)}\\
\ig[width=0.42\tew]{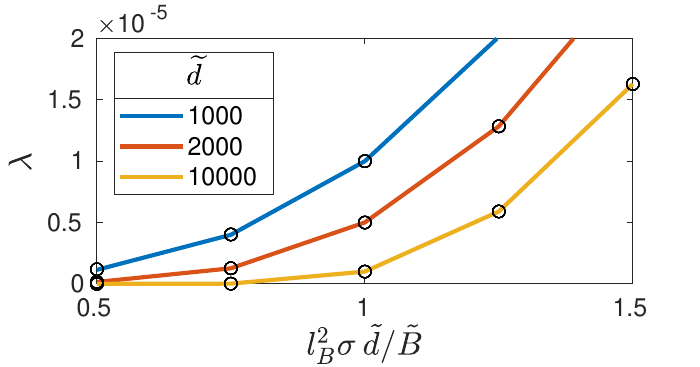}
}
\caption{{\sm (a,b) Profiles $\phi(\tl{x})$ and $n_+(x)/\lam_*$ for various values of
    $l_B^2\sigma$ and fixed $\tl{d}=10^3$ and
    $\tl{B}=10$. $\tl{x}=x/\l_B$ is the length in units of the Bjerrum
    length. (c) $\lam$ as a function of $l_B^2\sig$ for different $\tl{d}$ 
with fixed $\tl{B}/\tl{d}=10^{-2}$; full numerics for \reff{eq_shear} (lines) 
compared to numerical solution of \reff{eq_f_of_lam} ($\circ$).
\label{fig_phi_n}}}
\end{figure}

Fig.~\ref{fig_phi_n} shows quantitative results for \reff{eq_shear}, 
and an assessment of the approximation \reff{eq_f_of_lam}. To numerically solve \reff{eq_shear} 
for 
$\phi$ and $\lam$ 
we use the {\sc Matlab} solver {\tt bvp5c} with $\lam$ as part of the solution, 
and to solve \reff{eq_f_of_lam} we simply use {\tt fzero}.
In (a,b) 
we choose $(\tl{d},\tl{B})=(1000,10)$, hence $\lam_*=10^{-5}$, and vary 
$l_B^2\sig$. In (c) we plot $\lam$ as a function of $l_B^2\sig$ for different 
$\tl{d}$ values with fixed $\tl{B}/\tl{d}=10^{-2}$; the solid lines are from 
numerics, while the empty circles are from \reff{eq_f_of_lam}, with 
excellent agreement. For large $\tl{d}$, 
$\lam$ is essentially $0$ for $l_B^2\sig< d/B$, and altogether 
all results fully agree with those explained (semi-)analytically above.
\section{Electrolytes in Taylor-Couette flow}\label{sec_coax}
\begin{figure}[ht!]
  \centering
  \includegraphics[width=0.36\textwidth,bb=178 60 555 460,clip]{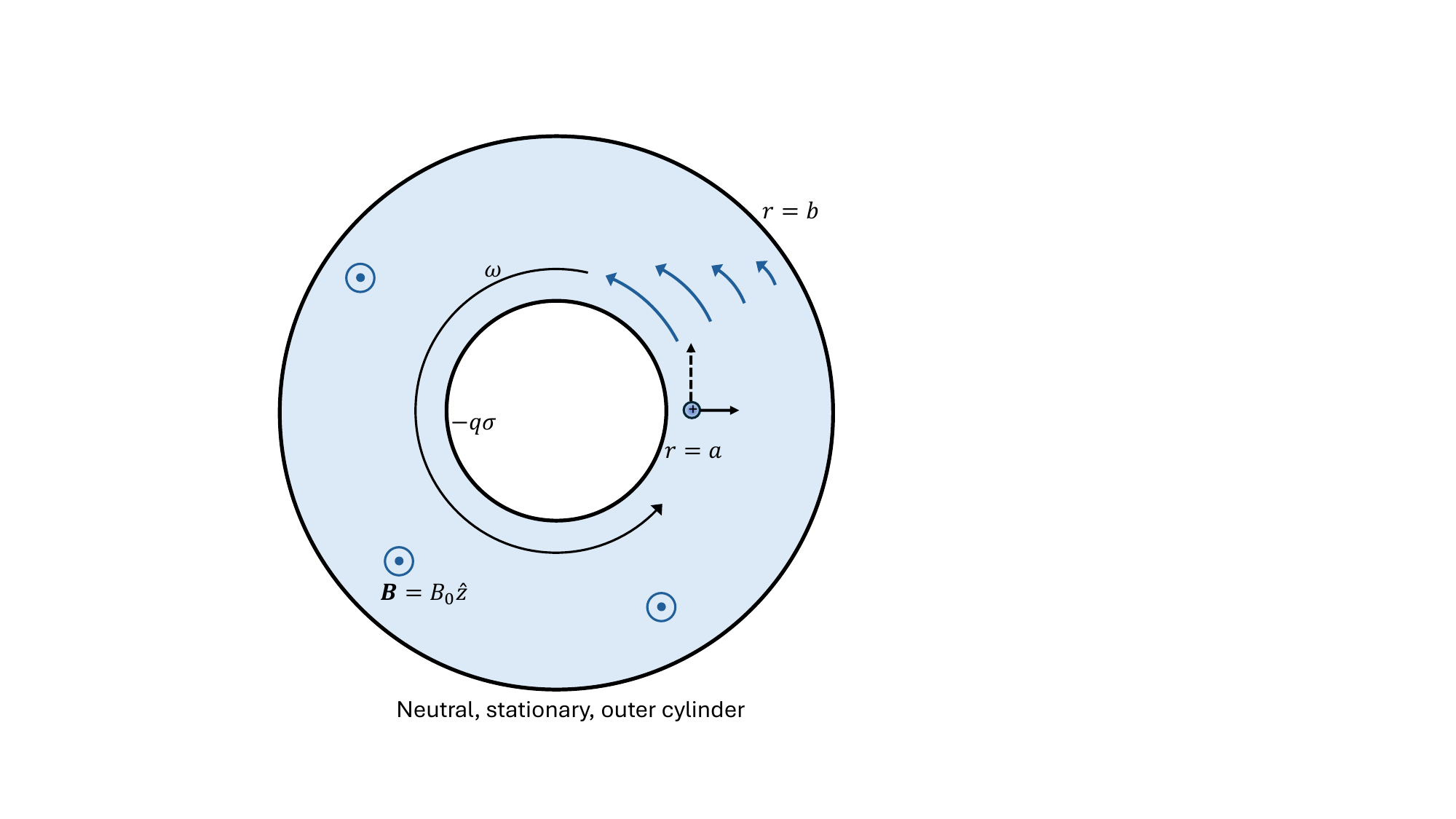}
  \caption{{\sm Illustration of electrolyte (light blue) with positive
    counterions confined between
    two coaxial cylinders with inner and outer radii $a$ and $b$,
    respectively. The inner cylinder has $-q\sigma$ surface charge
    density and rotates at a frequency $\omega$ while the outer one is
    stationary and electrically neutral. The magnetic field along the
    cylinders' axis ($z$ direction) leads to a Lorentz force in the
    radial direction.
}
  \label{fig_fig6_illus}}
\end{figure}

We have found counterion run-off in one Cartesian dimension. Let us
examine a two-dimensional system of two rotating cylinders and see
how the classical transition is affected here. Figure \ref{fig_fig6_illus} shows an electrolyte confined 
between two 
infinitely long
coaxial cylinders with radii $a$ and $b>a$. The inner one rotates at
an angular frequency $\omega$ and is charged with $-q\sigma$ Coulomb
per unit area ($\sigma$ is a surface number density). Its counterions
are released to the liquid. The outer cylinder is stationary and
neutral. The magnetic field is oriented along the cylinders' axis, 
${\bf B}=B_0\hat{z}$. Since ions move in the azimuthal direction, the
electric field acts in the radial direction, ${\bf E}=E\hat{r}$.

Let us look at the magnitude of centrifugal force compared to the
Lorentz force in the radial direction. The strongest acceleration
$v^2/r$ is at the inner cylinder, where $r=a$ and $v=\omega a$. Hence,
the centrifugal force acting on an ion is
$f_{\rm cent}=m_i\omega^2 a$, where $m_i$ is the ionic mass. Assuming a
Sodium ion ($23$ atomic mass units), $a=1$ cm and $\omega=100$
rad/s, we find $f_{\rm cent}\approx 4\times 10^{-24}$N. The
magnetic force for a unit charge moving at the same velocity ($1$ m/s)
in a large magnetic field of $0.1$ Tesla is $f_L=1.6\times
10^{-20}$N. With this estimate we safely neglect the effect of
centrifugal acceleration.
It is useful to look at the steady Navier-Stokes equations in polar coordinates $(r,\theta)$ 
\cite{bsl_book}, i.e.~for the $(r,\theta)$ components:
\begin{\eas}
  r:~~-\rho\frac{v_\theta^2}{r}&=&
  -\partial_rp+\rho E+jB_0-k_BT\partial_r(n^++n^-),\\
  \theta:~~~~~~~~0&=&\frac{1}{r^2}\partial_r\left(\eta
    r^3\partial_r\left(\frac{v_\theta}{r}\right)\right).
\end{\eas}
The total ionic flux ${\bf h}$ has two terms along $\hat{r}$ and advection
along $\hat{\theta}$:
\begin{\eas} {\bf h}&=&\left[n^+\mu q\left({\bf E}+{\bf v}\times{\bf
        B}\right)-D\nabla n^+\right]\hat{r}+n^+qv\hat{\theta}.
\end{\eas}
As in the Cartesian case, the ionic flux vanishes in the $r$
direction, and the $r$ equation simplifies to
$-\rho v_\theta^2/r=-\partial_r p$, yielding the pressure once
$v_\theta$ is known from the $\theta$ direction. The flow profile solving the $\theta$-equation is
\begin{\ea}\label{eq_v_coaxial} {\bf v}(r)&=&\frac{\omega
    a}{1-\left(\frac{a}{b}\right)^2}\frac{a}{b}\left(-\frac{r}{b}+\frac{b}{r}
  \right)\hat{\theta}.
\end{\ea}
The continuity equation for the ionic flux $\nabla\cdot{\bf h}=0$
means that
\begin{\eas}
  \frac{1}{r}\frac{\partial(r h_r)}{\partial
    r}+\frac{1}{r}\frac{\partial
    h_\theta}{\partial\theta}+\frac{\partial h_z}{\partial z}&=&0,\nn
\end{\eas}
and the cylindrical symmetry leads to
\begin{\eas}
  r\left[n^+\mu q\left({\bf E}+{\bf v}\times{\bf B}\right)-D\nabla
    n^+\right]&=&0,
\end{\eas}
where we used the vanishing flux at the hard walls at $r=a$ and
$r=b$. Integration of this equation gives the cation density as a
Boltzmann-like distribution
\begin{\eas}
  n^+(r)&=&n_m\exp\left(\frac{-q\psi+U_{\rm hyd}(r)}{k_BT}\right),
\end{\eas}
where $n_m$ is a yet-unknown constant and the hydrodynamic potential
is $U_{\rm hyd}(r)\equiv q \int_a^rv(r')B_0dr'$ with $v(r)$ given by 
(\ref{eq_v_coaxial}).
We now make a series of transformations: First, length is
measured in units of $a$, $\tl{r}=r/a$, and following Burak and Orland
\cite{burak_orland_pre_2006} we introduce the logarithmic coordinate
\huga{
\text{$u=\ln(\tl{r})$ with $0 \le u \le d:=\ln(b/a)$}
} 
to map the cylindrical domain to a rectangular one. We then define
\hugast{
  \phi=\frac{q\psi}{k_BT} -
  \frac{\tl{B}}{g}\left(-\frac{a}{2b}\tl{r}^2+
    \frac{b}{a}\ln(\tl{r}a/b)\right),
}
where
\hugast{
  v_0=\omega a,~~g=\frac{b}{a}-\frac{a}{b},~~
  \tl{B}=\frac{qv_0B_0a}{k_BT}.
}
Finally, we define $\vphi$ as $\vphi=\phi-2u$, removing the explicit
geometric and magnetic contributions. These steps yield the Poisson-Boltzmann equation 
\hual{
  \vphi''&=-\lambda e^{-\vphi}+2\frac{\tl{B}a}{gb}e^{2u}\label{eq_gov_eqn}, \\
  \vphi'(0)&=l_B\sigma a-2-\tl{B},\quad \vphi_u(d)=-2,\quad \vphi(0)=0.\nonu
}
As before, the last BC is a choice of the gauge, leading to a unique 
eigenvalue 
\huga{
\lambda=l_Bn_m'a^2 
}
for solving (\ref{eq_gov_eqn}), where 
\hugast{n'_m=n_m\exp\left(
\frac{\tl{B}}g\left(\frac a{2b}+\frac b a\ln(\frac b a)
\right)\right).
}
The term $2\tl{B}a/(gb)e^{2u}$ represents the inhomogeneous magnetic
forcing arising from the hydrodynamic potential in cylindrical
geometry. The  different definition of 
$\lambda$ (compared to  Sec.\ref{sec_shear}) arises from
the geometry: cylindrical vs planar. Similarly, $\tilde{B}$ is
divided by an additional geometric factor $g$ in cylindrical
coordinates,  and the boundary conditions in \reff{eq_gov_eqn} now yield 
the charge conservation 
\begin{\ea}
  \int_0^{d}\lambda e^{-\vphi}du&=&l_B\sigma a.
\end{\ea}

Before we continue, let us make a short recap of the regular
Manning-Oosawa (MO) condensation, recovered formally when $\tl{B}=0$,
occurring in the limit $\frac b a{\to}\infty$. Here, there are two scenarios:
when $l_B\sigma a<2$, the eigenvalue $\lam$ essentially vanishes, 
and the (approximate, with violation of the right BC) 
solution then is $\vphi_0=(l_B\sigma a-2)u$, i.e, 
$q\psi_0/k_BT=l_B\sigma a\ln(r/a)$, and the counterions ``run'' to
infinity. Since $n_m$, the concentration of counterions at the inner
cylinder, is proportional to $\lambda$, counterions' concentration
vanishes. Above the Manning criterion, $l_B\sigma a>2$,
$\vphi_0(u)=2\ln\left(1+\frac12(l_B\sigma a-2)u\right)$ (and
$\lambda=l_B\sigma a-2$). That is,
$q\psi_0/k_BT=2\ln(r/a)+2\ln\left[1+\frac12(l_B\sigma
  a-2)\ln(r/a)\right]$.
\iffalse 
\begin{figure}[ht]
  \centering
\btab{l}{{\sm (a)}\\
\ig[width=0.4\tew]{c1}\\
{\sm (b)}\\
\ig[width=0.4\tew]{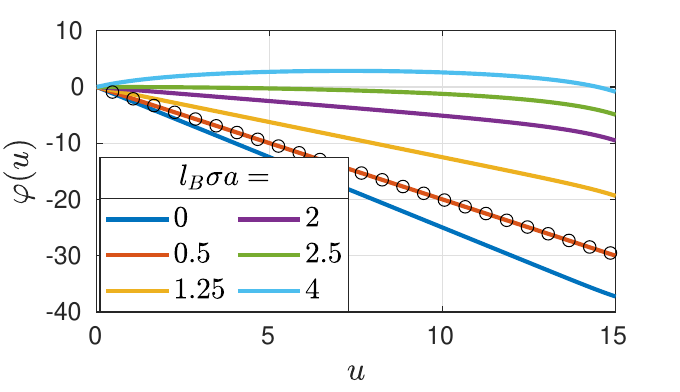}
}

\vs{-2mm}
  \caption{{\sm (a) $\lambda$ from (\ref{eq_gov_eqn}) vs dimensionless surface 
density of the
    inner cylinder.
Vertical dashed line marks the shifted Manning
    threshold $l_B\sigma a{=}2+\tl{B}$. Here $\ln(b/a){=}15$ and
    $\tl{B}=0.5$. (b) Solutions $\vphi(u)$ for selected values of
    $l_B\sigma a$ from (a). Blue curve for $\lambda=0$ is the approximation 
(\ref{eq_ana_lam0}) with $l_B\sigma a=0$, and circles are
    (\ref{eq_ana_lam0}) for the smallest non-zero value,
    $l_B\sigma a=0.5$.}
  \label{fig_lam_coax}} 
\end{figure}
\else 
\begin{figure}[ht]
  \centering
\btab{l}{{\sm (a)}\\
\ig[width=0.44\tew]{c2}\\
{\sm (b)}\\
\ig[width=0.44\tew]{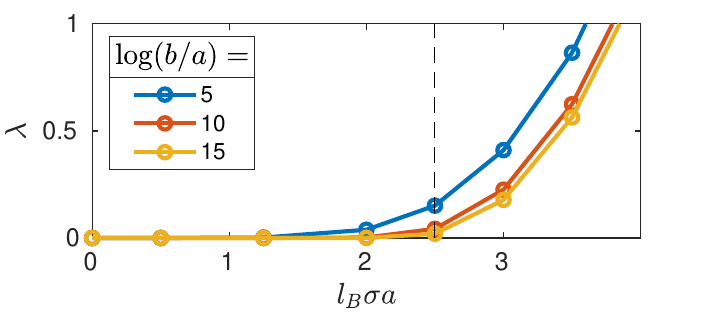}
}

\vs{-2mm}
  \caption{{\sm (a) Solutions $\vphi(u)$ for selected values of
    $l_B\sigma a$, $\log(b/a)=15$ and $\tl{B}=0.5$. Blue curve for $\lambda=0$ is the approximation 
(\ref{eq_ana_lam0}) with $l_B\sigma a=0$, and circles are
    (\ref{eq_ana_lam0}) for the smallest non-zero value,
    $l_B\sigma a=0.5$. (b) $\lambda$ from (\ref{eq_gov_eqn}) vs 
$l_B\sig a$ for different $\log(b/a)$, 
The vertical dashed line marks the shifted Manning
    threshold $l_B\sigma a{=}2+\tl{B}$. }
  \label{fig_lam_coax}} 
\end{figure}
\fi 

We now go back to the case with magnetic field. There is a transition
analogous to the MO condensation: In the limit $\frac b a\to\infty$,
$\lambda\approx 0$ when $l_B\sigma a<2+\tl{B}$. Counterions
escape radially outward and run off to infinity like in the MO
condensation. The MO criterion is shifted; the surface charge density
is renormalized such that $l_B\sigma a^*=l_B\sigma a-\tl{B}$. The
additional $\tl{B}$ term reflects the magnetic contribution to the
effective surface charge. There is also a quantitative difference due
to the inhomogeneous driving, but this is less essential to the
condensation transition.
For finite $b/a$ and $\lambda=0$, an approximate analytical solution to 
(\ref{eq_gov_eqn})
satisfying both $\vphi_u(0)=l_B\sigma a-2-\tl{B}$ and $\vphi(0)=0$ is 
\huga{\label{eq_ana_lam0}
  \vphi=\frac12\frac{\tl{B}a}{gb}e^{2u}+c_2u+c_1,
}
with $c_2=l_B\sigma
  a-2-\tl{B}\left(1+\frac{a}{gb}\right)$, and $c_1=-\frac12\frac{\tl{B}a}{gb}$.
The counterion concentration at the inner cylinder
vanishes, and ions migrate outward indefinitely, similar to the MO
runaway in cylindrical geometry. The mismatch of the right boundary
condition in this case is the same as in the MO case.
Figure \ref{fig_lam_coax}(a) shows selected profiles $\vphi(u)$, and 
comparison with \reff{eq_ana_lam0} 
for the smallest non-zero surface charge value, $l_B\sigma a=0.5$, giving 
excellent match with the numerical solution. Figure (b) 
shows the eigenvalue $\lambda$
calculated numerically from (\ref{eq_gov_eqn}) as a function of
$l_B\sigma a$. When $l_B\sigma a>2+\tl{B}$, $\lambda$ is positive and
$\vphi(u)$ is a concave function. When $l_B\sigma a$ decreases below
$2+\tl{B}$, $\lambda$ essentially vanishes for large $b/a$.\\[3mm]
\section{Conclusions}\label{dsec}
We presented a theoretical framework for electromagnetohydrodynamic (EMHD)
effects in electrolyte flows in a Stokes-Poisson-Nernst-Planck
formulation, focusing on how magnetic forcing interacts with
electrostatic screening to control ionic distributions. By introducing
a hydrodynamic potential that competes with electrostatics, we extend
classical MHD models beyond uniform conductivity and local
electroneutrality assumptions.
The analysis for unidirectional flows reveals novel mechanisms
for magnetic control of screening \cite{kornyshev_jpcb_2007} and 
identifies condensation-like transitions in driven systems.
Our results show that, even within the Stokes-Poisson-Nernst-Planck
framework, magnetic forcing can reconfigure ionic distributions
through a hydrodynamic potential.
This goes beyond standard MHD models, providing a
theoretical basis for magnetic control of screening in canonical flows
\cite{muller_book,bau_mech_res_comm_2009}.
We identify a non-electrostatic route to MO-like transitions,
including a condensation threshold in 1D (planar shear) and a shift of
the classical Manning criterion for cylindrical geometries. This
extends MO phenomenology to driven, non-equilibrium settings, while
retaining analytical tractability within a modified PB theory
\cite{manning_jcp_1969,oosawa_book,burak_orland_pre_2006,trizac_prl_2006,levin_physicaa_1998}.
For typical aqueous electrolytes at room temperature, the critical surface charge density required 
to 
observe the exact threshold is very low. However, the primary value of this work is establishing the 
fundamental existence of a non-geometric route to condensation. Furthermore, the transition is 
governed by the dimensionless parameter $\tl{B} = qv_0B_0d/k_BT$. This parameter can be 
enhanced 
by several orders of magnitude to reach experimentally accessible regimes in specific soft-matter 
environments. For instance, in highly viscous systems like ionic liquids or polymer melts, larger 
velocity gradients can be maintained without triggering turbulence. Similarly, the use of 
multivalent 
ions, lower temperatures, or microfluidic devices with extremely strong local magnetic fields can 
push 
the system toward the transition threshold.

\section*{Acknowledgments}

We are grateful for financial support by the Israel Science Foundation
Grant No. 332/24.

\bibliography{hmbib7}

\end{document}